\documentstyle[prb,aps,multicol,epsfig]{revtex}
\begin{document}
\draft
\title{Effective charges and statistical signatures in the noise
of normal metal--superconductor junctions at arbitrary bias}
\author{
Julien Torr\`es$^{a}$,
Thierry Martin$^{a}$
and Gordey~B.\ Lesovik$^{a, b}$
}
\address{
$^{a}$Centre de Physique Th\'eorique,
Universit\'e de la M\'editerran\'ee,
Luminy Case 907, F-13288 Marseille Cedex 9, France}
\address{
$^{b}$L.~D.~Landau Institute for Theoretical Physics, 117940 Moscow, Russia}
\maketitle

\begin{abstract}
Shot noise is studied in a single normal metal-superconductor (N-S)
junction at finite frequency,
and for branched N-S junctions at zero frequency. The noise
spectral density
displays a singularity at the Josephson frequency 
($\omega = 2eV/\hbar$) when
the applied bias is smaller than gap of the superconductor.
Yet, in the limit $eV\gg \Delta$, quasiparticle contributions
yield a singularity at $\omega = eV/\hbar$ analogous to that of
a normal metal. The crossover between these two regimes
shows new structures in the noise characteristic, 
pointing out the failure of the effective charge 
model. As an alternative to a finite frequency measurement, if a
sinusoidal external field is superposed to the constant bias
(non stationary Aharonov--Bohm effect), the second derivative of
the zero frequency noise with respect to the voltage exhibits
peaks when the frequency of the perturbation is commensurate
with the Josephson frequency. Finally, the statistical aspects
of noise are studied with an analog of the Hanbury-Brown and
Twiss experiment for fermions: a superconductor connected to two
normal leads. Noise correlations are found to be either negative
(fermionic) or positive (bosonic), due to the presence of
evanescent Cooper pairs in the normal side of the junction, in
the latter case. \end{abstract} 
\begin{multicols}{2}
%
\section{Introduction}
Recent developments in condensed matter physics emphasize
the importance of shot noise
in mesoscopic conductors. Noise contains
more information than the conductance:
in the Poisson regime for instance,
zero-frequency noise is proportional to the current
and to the effective charge. 
A convincing application is the direct measurement of the
fractional charge in a quantum Hall liquid \cite{Saminadayar
Reznikov}.
Another important issue of noise concerns the statistics of the 
(quasi-) particles. It is now established that Hanbury-Brown
and Twiss \cite{HBT} type correlation experiments yield a 
different sign for bosons and
for fermions
\cite{
Martin Landauer,Buttiker,Henny,Oliver}.

This paper focuses on both issues in
normal metal-superconductor (NS) junctions.
Existing results on junctions between two normal metals (NN)
are summarized below.
Finite frequency shot noise has a singularity at 
$\hbar \omega = eV$ (where $V$ is the applied bias) \cite{Yang},
which was detected experimentally \cite{Schoelkopf1}.
Another phenomenon, analogous to a finite frequency measurement, 
called the ``Non-Stationary Aharonov--Bohm effect'' 
\cite{Lesovik Levitov} uses a local alternating field superposed
to the bias voltage. Steps in the noise derivative with respect
to the DC bias were predicted \cite{Lesovik Levitov}
and measured \cite{Schoelkopf2}. The height of these
steps is non-monotonic with the amplitude of the harmonic 
perturbation. 

The statistical signatures of noise correlations
can be illustrated in a three terminal device where
current fluctuations in the two receiving leads are expected
to be fully anti-correlated \cite{Martin Landauer,Buttiker},
a direct consequence of the Pauli principle.
Experiments were performed in the integer
quantum Hall regime \cite{Henny} and in a two dimensional
electron gas \cite{Oliver}, with a splitter used to partition an
incident beam of electrons into reflected and transmitted
beams. The measurement of the correlations between the
reflected and the transmitted beams reported a negative value,
confirming the theoretical predictions.

In contrast to normal junctions, transport in NS junctions 
involves Andreev reflection \cite{Andreev}: an incoming 
electron is reflected as a hole at the boundary. The remaining
charge $2e$ is absorbed by the superconductor via a Cooper pair.
In a previous paper\cite{Lesovik Martin Torres}, finite
frequency noise was computed in the Andreev regime only ($eV \ll
\Delta$). The noise characteristics then present a singularity
at the frequency $\omega=2eV/\hbar$, a signature of an effective
charge $2e$ corresponding to that of a Cooper pair.
This issue is not surprising, because in the Andreev regime, 
some physical quantities can, in principle, be extracted from
the normal metal results by replacing $e$ by $2e$.
In particular, at zero-frequency the doubling of shot noise and
the crossover from thermal noise to excess noise have been
predicted \cite{Khlus,de Jong,Martin} and recently measured
\cite{Sanquer,Kozhevnikov}. However, for thermal noise (at
$k_BT\ll \Delta$), this naive substitution would imply a
violation of the fluctuation dissipation theorem.

Failures of this effective charge model also occur 
when the bias is comparable to the gap or
when it is increased beyond it. Indeed, when $eV>\Delta$,
the effective charge transfer picture breaks down, as quasi-particles
above the gap bear a single electron charge. It will be shown
that at such voltages, the
cusps/singularities which appear in the finite frequency
noise neither correspond to a charge $e$ nor to a charge $2e$.
Thus, a careful analysis and an explicit calculation 
are especially needed to describe the
crossover from sub-gap to above gap regime. Moreover, 
with the advent of superconducting samples with a ``small''
gap, this also allows to consider the limit
of large biases ($eV \gg \Delta$), where single quasiparticle
transmission overrides Andreev reflection. 
In this situation, one recovers the results obtained
for the NN junctions, with a charge transfer $e$.

At the same time, an NS junction which contains a splitter on
the normal side constitutes a Hanbury-Brown and Twiss 
type experiment where statistical effects can be detected
\cite{Torres Martin}. 
In the Andreev regime, the noise correlations
could possibly be positive (bosonic) due to the presence of 
Cooper pairs on the normal side (proximity effect), 
whereas a quasi-particle  dominated regime should 
favor negative (fermionic) correlations. 
The purpose is here to investigate the effective charges 
and statistical tendencies which show up in
the noise  characteristics of an NS junction in an united
fashion. Results  
for the sub-gap (Andreev) regime will be recalled,
and confronted to novel results for above gap transport. 

The paper is organized as follows. Noise correlations
in a multi-terminal device coupled with a
superconductor are computed at finite frequencies
(section \ref{para_noise_expr}). 
In section \ref{para_singleNS} a single N-S junction is studied
for several cases: a) in the Andreev regime (section
\ref{para_small_biases}), the noise spectral density presents a
singularity at the Josephson frequency; b) when the applied bias
is increased beyond the gap (section \ref{para_large_bias}),
additional singularities appear; c) if a sinusoidal external
field is added (non-stationary Aharonov-Bohm effect, section
\ref{para_NSAB}), the second derivative of noise with respect to
the bias presents peaks when the frequency of the perturbation is
commensurate with the Josephson frequency.
The last section (\ref{para_HBT}) deals with the
fermionic Hanbury-Brown and Twiss experimental proposal
with a superconductor, showing that noise correlations
may be either negative or positive.
\section{Current, noise and noise correlations}
\label{para_noise_expr}
\subsection{Assumptions}
The superconductor is connected to an arbitrary
number of normal leads (multi-terminal device),  
and the device is supposed to be small enough
that all scattering processes are elastic.
Note that the case of two superconductors
is not addressed here (see Ref. \cite{Anantram Datta}).
Calculations are restricted to the one-channel case, 
but a generalization to a multi-channel system is
straightforward. Thus, transport is dominated by the scattering
properties of the N-S junction \cite{Landauer,BILP,Imry,Datta}.
The two scattering processes at play are then 
Andreev reflection and normal reflection
of electrons and holes. 
The superconductor is maintained at a constant chemical
potential $\mu_S$, and each normal terminal is fixed at 
the same potential $\mu_{N}$.
For convenience, all energies are measured with respect to 
$\mu_S$, so that the applied bias reads 
$\mu_N-\mu_S=eV$. This bias is chosen to be positive throughout
the paper. 

Shot noise in a given lead, or alternatively noise
correlations between two (normal) terminals  
are defined as the Fourier transform of the current-current
correlation function:
\begin{eqnarray}
\nonumber
S_{i j}(\omega)
&=&
\lim_{T \rightarrow +\infty} \frac{1}{T}
\int_{-T/2}^{T/2} dt \int_{-\infty}^{+\infty} dt' e^{i \omega t'}
\\
\label{eq_noise_def}
&& \hspace{1.5cm}
\Bigl(
\langle I_i(t) I_j(t+t') \rangle
- \langle I_i \rangle \langle I_j \rangle
\Bigr)
~.
\end{eqnarray}
When $i=j$, $S_{ii}(\omega)$
corresponds to the noise in the terminal $i$,
whereas if $i$ and $j$ are different,
$S_{i j}(0)$ 
represents the zero frequency noise correlations between
lead $i$ and lead $j$.
$\langle ~ \rangle$ designates the thermodynamical average
in the grand-canonical ensemble.
\subsection{Bogolubov-de Gennes equations}
The Bogolubov-de Gennes \cite{de Gennes} (BdG) approach to
inhomogeneous  superconductivity is the adapted 
formalism to treat
electrons and holes on the same footing.
Performing the Bogolubov transformation (which
must diagonalize the effective BdG Hamiltonian),
and going to an energy representation,  
the annihilation operator of a particle with spin $\sigma$
($\sigma = \pm 1$) at the position $x$
in the terminal $i$ $\psi_{i, \sigma}(x)$
can be written as:
\begin{eqnarray}
\nonumber
\psi_{i, \sigma}(x) &=& \frac{1}{\sqrt{2 \pi}} \sum_j \sum_{\beta} \int_0^{+\infty} d E
\left( \frac{u_{i \, j \, \beta}(x)}{\sqrt{\hbar v^j_e(E)}} c_{j \, \beta \, \sigma}(E) \right.
\\
\label{eq_bog_trans}
&& \hspace{1.5cm}
\left. - \sigma \frac{v_{i \, j\, \beta}^*(x)}{\sqrt{\hbar v^j_h(E)}} c^{+}_{j \, \beta \, -\sigma}(E) \right)
~.
\end{eqnarray}
State $u_{i \, j \, \beta}$ ($v_{i \, j \, \beta}$) corresponds
to the wave function of a  electron (a hole) scattered in terminal
$i$, due to a quasi-particle of type $\beta$ 
(electron or hole, $\beta=e,h$) which was incoming
from lead $j$. Operators $c(E)$ and $c^{+}(E)$ 
satisfy standard anticommutation
relations. $v^j_e(E)=\hbar k_e^j(E)$ is the
velocity in the lead $j$. The Bogolubov-de Gennes equations
may be written as:
\begin{equation}
\label{eq_BdG}
\left\{
\begin{array}{llll}
E u_{i \, j\, \beta}(x) &=&   & \!\!\! \left( -\displaystyle \frac{\hbar^2}{2m}\frac{\partial^2}{\partial x^2} - \mu_S + V(x) \right) u_{i \, j\, \beta}(x)
\\
\\
&&& \hspace{2cm} + \Delta(x) v_{i \, j\, \beta}(x) ~,
\\
\\
E v_{i j \beta}(x) &=& - & \!\!\! \left( -\displaystyle \frac{\hbar^2}{2m}\frac{\partial^2}{\partial x^2} - \mu_S + V(x) \right) v_{i \, j\, \beta}(x)
\\
\\
&&& \hspace{2cm} + \Delta^*(x) u_{i \, j\, \beta}(x)
~,
\end{array}
\right.
\end{equation}
and describe the evolution of particle states $u_{i \, j\, \beta}$
and $v_{i \, j\, \beta}$. The pair potential
$\Delta(x)$ should be calculated self-consistently, but
for simplicity corresponds here to the superconducting gap 
in the bulk superconductor ($x>0$), and gives zero in the normal
terminals ($x<0$).
\subsection{States in a normal terminal}
\label{para_etats_normal}
In order to calculate noise with Eq. (\ref{eq_noise_def}),
the states which appear in the Bogolubov transformation
(\ref{eq_bog_trans}) are specified using the scattering 
($S$) matrix describing the junction. 
In a normal, ideal  lead $\Delta(x)=0$ and $V(x)=0$, the
Bogolubov-de Gennes equations (\ref{eq_BdG}) reduces to a
Schr\"odinger equation for electrons, and to its time reversed
analog for holes.
Solutions of the form $e^{ik^N_e x}$
for electrons, and $e^{ik^N_h x}$ for holes are chosen,
where $k^N_e=\sqrt{2m \left( \mu_S + E \right)}/\hbar$ 
and $k^N_h= \sqrt{2m \left( \mu_S - E \right)}/\hbar$ 
are the wave vectors of electrons and holes.
Electrons and holes in lead $i$ which originate from
a particle of type $\beta$ ($\beta=e,h$) in terminal $j$
and scattered into $i$ are described by:
\begin{eqnarray}
u_{i \, j\, \beta} (x) &=& \delta_{i,j} \delta_{e,\beta}
e^{ik^N_e x} + s_{ i j e \beta}
\sqrt{\frac{k^j_\beta}{k^N_e}} e^{-ik^N_e x} ~,
\label{u in terms of s}
\\
v_{i \, j\, \beta} (x) &=& \delta_{i,j} \delta_{h,\beta}
e^{-ik^N_h x} + s_{ i j h \beta}
\sqrt{\frac{k^j_\beta}{k^N_h}} e^{ik^N_h x} ~.
\label{v in terms of s}
\end{eqnarray} 
Note the opposite sign for momenta of electrons and holes.
$x_i$, the position in terminal $i$ is 
specified as $x$ in Eqs. (\ref{u in terms of s}) and 
(\ref{v in terms of s}) for simplicity of notation. $s_{ i j \alpha \beta}$
is the scattering matrix element expressing the
amplitude of an outgoing particle $\alpha$ in lead $i$
due to an incident particle of type $\beta$ in lead $j$.
\subsection{Current operator and average current}
The current operator in lead $i$ is defined by:
\begin{eqnarray}
\nonumber
I_i(x) &=&
e \frac{\hbar}{2mi} \sum_{\sigma}
\left(
  \psi^+_{i, \sigma}(x) \frac{\partial \psi_{i, \sigma}(x)}{\partial x}
\right.
\\
&& \hspace{2cm}
\left.
- \frac{\partial \psi^+_{i, \sigma}(x)}{\partial x} \psi_{i, \sigma}(x)
\right)
~.
\end{eqnarray}
Substituting $\psi$ by Eq. (\ref{eq_bog_trans}),
the current operator becomes:
\begin{eqnarray}
\nonumber
I_i(x) &=&
\frac{e \hbar}{2mi v_F} \frac{1}{2 \pi \hbar} \int_0^{+\infty} d E_1\int_0^{+\infty}  dE_2 \sum_{m,n} \sum_{\sigma}
\\
\nonumber
&& \hspace{0.9cm}
\Bigl[
\Bigl(
u_{i m}^* \partial_x u_{i n} - \partial_x u_{i m}^* u_{i n}
\Bigr) c^+_{m\,\sigma} c_{n\,\sigma}
\\
\nonumber
&& \hspace{0.9cm}
- \Bigl(
u_{i m}^* \partial_x v_{i n}^* - \partial_x u_{i m}^* v_{i n}^*
\Bigr) \sigma \, c^+_{m\,\sigma} c^+_{n\,-\sigma}
\\
\nonumber
&& \hspace{0.9cm}
- \Bigl(
v_{i m} \partial_x u_{i n} - \partial_x v_{i m} u_{i n}
\Bigr) \sigma \, c_{m\,-\sigma} c_{n\,\sigma}
\\
\label{eq_current}
&& \hspace{0.9cm}
+\Bigl(
v_{i m} \partial_x v_{i n}^* - \partial_x v_{i m} v_{i n}^*
\Bigr) c_{m\,-\sigma} c^+_{n\,-\sigma}
\Bigr]
~,
\end{eqnarray}
where the sums over $j$ and $\beta$ have been replaced
by a single sum over index $m$.
Expressions with index  $m$ ($n$) have an energy dependence
$E_1$ ($E_2$).
Solving the Bogolubov-de Gennes equations in the superconductor leads to find
the waves vectors
$k^S_{e,h}
 = \sqrt{2m} \left(\mu_S \pm \left( E^2 - \Delta^2 \right)^{1/2}\right)^{1/2}/\hbar$
for electron-like (+) and hole-like (-) quasi-particles.
The chemical potential
of the superconductor $\mu_S$ is large compared to all the energy
scales involved in the system. Thus, the assumption
$k^N_e=k^N_h=k^S_e=k^S_h=k_F$ is made, which
explains the presence of the Fermi velocity $v_F$
in Eq. (\ref{eq_current}). 

The calculation of the average current involves
the average of creation and annihilation operators
products such as $\langle c^+_{m\,\sigma}(E_1)
c_{n\,\sigma}(E_2) \rangle = f_m(E_1) \delta_{mn}
\delta(E_1-E_2)$.
$f_m(E)=f_{FD}(E\mp eV)$ for electrons and holes
on the normal side, with  $f_{FD}(E)$ the Fermi-Dirac
distribution. $f_m(E)=f_{FD}(E)$ in the superconductor. 
The average current is then:
\begin{eqnarray}
\nonumber
\langle I_i(x) \rangle &=&
\frac{e}{2\pi mi v_F} \int_0^{+\infty} d E \sum_{m}
\\
\nonumber
&& \hspace{0.5cm}
\Biggl[
\Bigl(
u_{i m}^* \partial_x u_{i m} - \partial_x u_{i m}^* u_{i m}
\Bigr) f_m
\\
&& \hspace{0.5cm}
+\Bigl(
v_{i m} \partial_x v_{i m}^* - \partial_x v_{i m} v_{i m}^*
\Bigr) \left(1-f_m\right)
\Biggr]
~.
\end{eqnarray}
\subsection{Noise and noise correlations}
Since a single current operator is composed of
products of two creators or annihilators, 
$\langle I_i(x,t) I_j(x,t+t') \rangle$
is a sum of average values of four creation or annihilation
operators.
These average values are expressed as a function of
the Fermi-Dirac distributions using Wick's theorem.
The calculation of noise is now performed using Eq. (\ref{eq_noise_def}).
It is convenient to define the following matrix elements:
\begin{eqnarray}
\nonumber
A_{i m j n}(E,E',t) &=&
u_{j n}(E',t) \partial_x u^*_{i m}(E,t)
\\
&&
 -  u^*_{i m}(E,t) \partial_x u_{j n}(E',t) ~,
\\
\nonumber
B_{i m j n}(E,E',t) &=&
v^*_{j n}(E',t) \partial_x v_{i m}(E,t)
\\
&&
- v_{i m}(E,t) \partial_x v^*_{j n}(E',t) ~,
\\
\nonumber
C_{i m j n}(E,E',t) &=&
u_{j n}(E',t) \partial_x v_{i m}(E,t)
\\
&&
- v_{i m}(E,t) \partial_x u_{j n}(E',t)
~.
\end{eqnarray}
Calculating all the average values and the difference
$\langle I_i(t) I_j(t+t') \rangle
- \langle I_i \rangle \langle I_j \rangle$,
one obtains the noise or the noise correlations:
\begin{eqnarray}
\nonumber
S_{ij}(\omega) &=&
\frac{e^2 \hbar^2}{2m^2 v_F^2} \frac{1}{\left( 2 \pi \hbar \right)^2}
\lim_{T\rightarrow +\infty} \frac{1}{T} \int_{-T/2}^{+T/2} \!\!\! dt \int_{-\infty}^{+\infty} \!\!\! dt' e^{i\omega t'}
\\
\nonumber
&& \hspace{5mm} \times
\int_0^{+\infty} \!\!\! dE \int_0^{+\infty} \!\!\! dE' \sum_{m,n} \Biggl\{
\\
\nonumber
&& \hspace{5mm}
f_m(E)(1-f_n(E')) e^{i(E'-E)t'/\hbar}
\\
\nonumber
&& \hspace{5mm}
\times \Bigl[
  A_{i m j n}(E,E',t)   A^*_{i m j n}(E,E',t+t')
\\
\nonumber
&& \hspace{5mm}
+ B^*_{i m j n}(E,E',t) B_{i m j n}(E,E',t+t')
\\
\nonumber
&& \hspace{5mm}
+ A_{i m j n}(E,E',t)   B_{i m j n}(E,E',t+t')
\\
\nonumber
&& \hspace{5mm}
+ B^*_{i m j n}(E,E',t) A^*_{i m j n}(E,E',t+t')
\Bigr]
\\
\nonumber
&& \hspace{5mm}
+f_m(E) f_n(E') e^{-i(E+E')t'/\hbar}
\, C^*_{i m j n}(E,E') 
\\
\nonumber
&& \hspace{1cm}
\times
\Bigl( C_{\j n i m}(E',E,t+t')
\\
\nonumber
&& \hspace{2cm}
 + C_{i m j n}(E,E',t+t') \Bigr)
\\
\nonumber
&& \hspace{5mm}
+(1-f_m(E)) (1-f_n(E')) e^{i(E+E')t'/\hbar}
\\
\nonumber
&& \hspace{5mm}
\times  \Bigl( C_{\j n i m}(E',E,t) + C_{i m j n}(E,E',t) \Bigr)
\\
\label{eq_corr_temp}
&& \hspace{1cm}
\times
C^*_{i m j n}(E,E',t+t')
\Biggr\}
~.
\end{eqnarray}
So far as the process is stationary (for example if
no time dependent external field is applied), matrix elements
$A_{i m j n}$, $B_{i m j n}$ et $C_{i m j n}$
are time independent. As a consequence, the integration over $t$
simplifies, and the integration over $t'$ gives $\delta$
functions with energy. As a result, terms proportional to 
$(1-f_m)(1-f_n)$ cancels because one is dealing with
quasi-particles with a {\it positive} energy. This yields:
\begin{eqnarray}
\nonumber
S_{ij}(\omega) &=&
\frac{e^2 \hbar^2}{2m^2 v_F^2} \frac{1}{ 2 \pi \hbar }
\int_0^{+\infty} \!\!\! dE \sum_{m,n} \Biggl\{
\\
\nonumber
&& \hspace{2mm}
   \Theta(E + \hbar \omega) f_m(E + \hbar \omega)(1-f_n(E))
\\
\nonumber
&& \hspace{3mm}
\times \bigl|
  A_{i m j n}(E + \hbar \omega,E) + B^*_{i m j n}(E + \hbar \omega,E)
\bigr|^2
\\
\nonumber
&& \hspace{2mm}
+ \Theta(\hbar \omega - E) f_m(\hbar \omega - E) f_n(E)
\\
\nonumber
&& \hspace{1cm}
\times C^*_{i m j n}(\hbar \omega - E,E,t) 
\\
\nonumber
&& \hspace{1cm}
\times
\Bigl( C_{\j n i m}(E,\hbar \omega - E)
\\
\nonumber
&& \hspace{2cm}
+ C_{i m j n}(\hbar \omega - E,E) \Bigr)
\\
\nonumber
&& \hspace{2mm}
+ \Theta( - E - \hbar \omega) (1-\!f_m( - E - \hbar \omega)) (1-\!f_n(E))
\\
\nonumber
&& \hspace{0.5cm}
\times C^*_{i m j n}( - E - \hbar \omega,E)
\\
\nonumber
&& \hspace{0.5cm}
\times
\Bigl( C_{\j n i m}(E, - E - \hbar \omega)
\\
&& \hspace{2cm}
+ C_{i m j n}( - E - \hbar \omega,E) \Bigr)
\Biggr\}
~,
\end{eqnarray}
after integration over $E'$.
This expression corresponds to the noise if
indices $i$ and $j$ are the same:
\begin{eqnarray}
\nonumber
S_{ii}(\omega) &=&
\frac{e^2 \hbar^2}{2m^2 v_F^2} \frac{1}{ 2 \pi \hbar }
\int_0^{+\infty} \!\!\! dE \sum_{m,n} \Biggl\{
\\
\nonumber
&& \hspace{2mm}
   \Theta(E + \hbar \omega) f_m(E + \hbar \omega)(1-f_n(E))
\\
\nonumber
&& \hspace{5mm}
\times \bigl|
  A_{i m i n}(E + \hbar \omega,E) + B^*_{i m i n}(E + \hbar \omega,E)
\bigr|^2
\\
\nonumber
&& \hspace{2mm}
+ \Theta(\hbar \omega - E) f_m(\hbar \omega - E) f_n(E)
\\
\nonumber
&& \hspace{5mm}
\times C^*_{i m i n}(\hbar \omega - E,E,t) 
\\
\nonumber
&& \hspace{5mm}
\times
\Bigl( C_{i n i m}(E,\hbar \omega - E) + C_{i m i n}(\hbar \omega - E,E) \Bigr)
\\
\nonumber
&& \hspace{2mm}
+ \Theta( -E - \hbar \omega) (1-f_m(-E - \hbar \omega)) (1-f_n(E))
\\
\nonumber
&& \hspace{5mm}
\times C^*_{i m i n}(-E - \hbar \omega,E)
\\
\nonumber
&& \hspace{5mm}
\times
  \Bigl( C_{i n i m}(E,-E - \hbar \omega)
\\
\label{eq_bruit}
&& \hspace{2cm}
+ C_{i m i n}(-E - \hbar \omega,E) \Bigr)
\Biggr\}
~.
\end{eqnarray}
The zero-frequency limit for noise correlations between terminals
$i$ and $j$ yields \cite{Martin,Anantram Datta,Datta Bagwell Anantram,Blanter_Buttiker}:
\begin{eqnarray}
\nonumber
S_{ij}(0) &=&
\frac{e^2 \hbar^2}{2m^2 v_F^2} \frac{1}{ 2 \pi \hbar }
\int_0^{+\infty} \!\!\! dE \sum_{m,n}  f_m(E)(1-f_n(E))
\\
\label{eq_corr_bruit}
&& \hspace{1cm}
\times
\bigl|  A_{i m j n}(E,E)  + B^*_{i m j n}(E,E) \bigr|^2
~.
\end{eqnarray}
\section{Single N-S junction}
\label{para_singleNS}
\subsection{General expression for noise}
A single N-S junction with arbitrary transparency (with a
tunneling barrier at the interface) is considered first for a
small applied bias (Andreev regime $eV \ll \Delta$) and then for
biases larger than the gap. Analytical expressions are
obtained in the first case, while numerical results will be
presented in the latter regime. For simplicity, the temperature
is chosen to be much smaller than the gap in both cases. 
In the the Andreev regime, the integrals over energy
in Eq. (\ref{eq_bruit}) can be  performed, resulting in three
distinct contributions combining different products of
Fermi-Dirac distributions. It is interesting to note
that although the current operator of Eq. (\ref{eq_current}) 
itself cannot couple states which differ by two 
quasi-particles, its fluctuations give a contribution which is
proportional to $f_n f_m$  in this finite frequency
calculation.
Rewriting down $A_{mn}$, $B_{mn}$ and
$C_{mn}$ as a function of the $S$ matrix elements, three
distinct expressions of the noise are found. If $\hbar \omega < eV$:
\begin{eqnarray}
\nonumber
S(\omega) & = & \frac{2 e^2}{h} \Biggl\{ \int_{ 0}^{eV - \hbar \omega} dE
\\
\nonumber
&& \hspace{2mm}
\times \Bigl[
|s_{NNee} (E+\hbar \omega)|^2 \left( 1-|s_{NNee}(E)|^2 \right) 
\\
\nonumber
&& \hspace{6mm}
+ |s_{NNhe} (E+\hbar \omega)|^2 \left( 1-|s_{NNhe}(E)|^2 \right)
\\
\nonumber
&& \hspace{6mm}
+ s_{NNee}^*(E+\hbar \omega) \; s_{NNee}(E)
\\
\nonumber
&& \hspace{9mm}
 \times \; s_{NNhe}(E+\hbar \omega) \; s_{NNhe}^*(E)
\\
\nonumber
&& \hspace{6mm}
+ s_{NNee}(E+\hbar \omega) \; s_{NNee}^*(E)
\\
\nonumber
&& \hspace{9mm}
 \times \; s_{NNhe}^*(E+\hbar \omega) \; s_{NNhe}(E)
\Bigr]
\\
\nonumber
&& \hspace{1mm}
+ \int_{ 0}^{\hbar \omega} dE \Bigl[
|s_{NNee}(E)|^2 |s_{NNhe} (\hbar \omega-E)|^2
\\
\nonumber
&& \hspace{3mm}
+ s_{NNee}(\hbar \omega-E) \; s_{NNhe}(E)
\\
\label{eq_int_bruit_1}
&& \hspace{9mm}
\times \; s_{NNee}^*(E) \; s_{NNhe}^*(\hbar \omega-E)
\Bigr]
\Biggr\}
~.
\end{eqnarray}
If $eV < \hbar \omega < 2 eV$, one obtains:
\begin{eqnarray}
\nonumber
S(\omega) & = & \frac{2 e^2}{h} \int_{\hbar \omega - eV}^{eV} dE \Bigl[
|s_{NNee} (E) |^2 |s_{NNhe} (\hbar \omega - E) |^2
\\
\nonumber
&& \hspace{1.7cm}
+ s_{NNee} (\hbar \omega - E) \; s_{NNhe} (E)
\\
\label{eq_int_bruit_2}
&& \hspace{1.9cm}
\times s^*_{NNee} (E) \; s^*_{NNhe} (\hbar \omega - E)
\Bigr]
~.
\end{eqnarray}
If $\hbar \omega > 2 eV$ noise vanishes.
Eq. (\ref{eq_int_bruit_2}) contains no spatial dependence as a
consequence of the approximation $\mu_S\gg\hbar\omega,eV$
mentioned in Sec. (\ref{para_etats_normal}). 

\subsection{Small biases: Andreev regime}  
\label{para_small_biases}
\begin{figure}[htb]
\begin{center}
\mbox{\epsfig{file=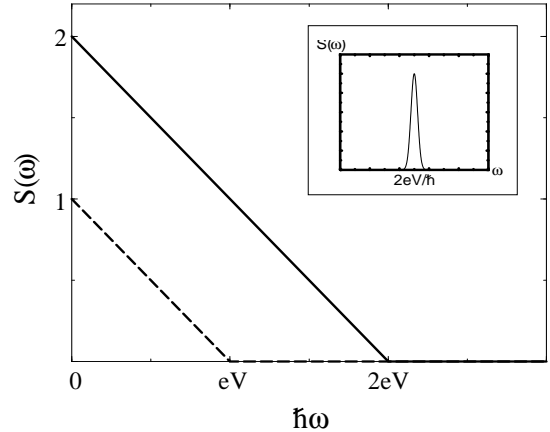,width=8.5cm}} \narrowtext
\caption[to]{ Noise as a function of frequency.
Full line: N-S junction, with a singularity
at $\hbar \omega = 2eV$ (in units of
$(4e^2/h)eV R_A(1-R_A)$); dashed line: junction between two
normal metals with a singularity at $\hbar \omega = eV$
(in units of $(4e^2/h)eV T(1-T)$). Inset:
noise in the Josephson effect, with a peak at the frequency
$\hbar \omega = 2eV$.
}
\label{bruitsing}
\end{center}
\end{figure}

When the applied bias is much smaller than the gap, 
the $S$ matrix elements can be taken to be constant. Using
the unitarity of the $S$ matrix, both expressions
(\ref{eq_int_bruit_1}) and (\ref{eq_int_bruit_2}) are
unified in the same formula 
over the whole energy interval $0 < \hbar \omega < 2 eV$:
\begin{equation}
\label{eq_bruit_NS_simple}
\left\{
\begin{array}{llll}
S(\omega) & = & \displaystyle \frac{4 e^2}{h} (2 eV - \hbar \omega) R_A(1-R_A) & \mbox{if $\hbar \omega < 2 eV$} ~,
\\ \\
S(\omega) & = & 0 & \mbox{if $\hbar \omega > 2 eV$} ~,
\end{array}
\right.
\end{equation}
where $R_A=|s_{NNhe}(0)|^2$ is the Andreev reflection probability.
The noise spectral density 
decreases linearly with frequency, and vanishes beyond 
the Josephson frequency $2 eV/\hbar$ (figure \ref{bruitsing}), 
thus displaying a singularity at this frequency.

\begin{figure}[htb]
\begin{center}
\mbox{\epsfig{file=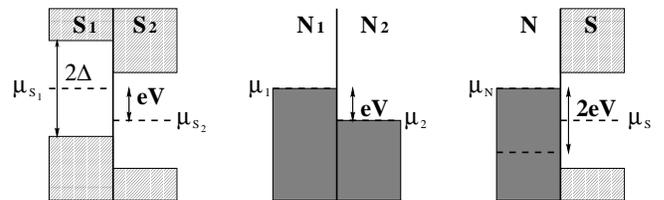,width=8.5cm}}
\narrowtext
\caption[to]{
Three different kinds of junctions:
a) two superconductors in contact (Josephson effect), with
chemical potentials $\mu_{S_1}$ and $\mu_{S_2}$;
b) two normal metals with
chemical potentials $\mu_1$ and $\mu_2$;
c) a normal metal ($\mu_N$) connected to a superconductor ($\mu_S$).
}
\label{sn}
\end{center}
\end{figure}

This result has to be compared with both the Josephson effect 
\cite{Josephson} and with the analog result for a normal
metal junction \cite{Yang} (figure
\ref{bruitsing}). In the former case, a DC bias applied to a 
junction between two superconductors generates an 
oscillatory current.
The order parameter on each side oscillates
as $\psi_{1,2} \sim \exp[-i2 \mu_{S_{1,2}}t/\hbar]$ with
$\mu_{S_1}$ and $\mu_{S_2}$ the chemical potentials of each
superconductor (figure \ref{sn}a). The resulting current
involves the overlap of these two states $\psi_1 \psi_2^*$, and
therefore oscillates at the frequency
$2|\mu_{S_2}-\mu_{S_1}|/\hbar$. The noise
characteristic exhibits a peak at $2eV/\hbar$ which radiation
line-width was computed in Ref. \cite{Larkin Ovchinnikov} (inset
figure \ref{bruitsing}).  In the former case (figure \ref{sn}b),
the wave functions have a time dependence as $\psi_{1,2} \sim
\exp[-i \mu_{1,2}t/\hbar]$, so that although the resulting
current is constant, finite frequency noise involves the overlap
$\psi_1 \psi_2^*$ leading to a singularity at the frequency
$|\mu_2-\mu_1|/\hbar=eV/\hbar$. 

In the N-S case (figure
\ref{sn}c), only Andreev reflection contributes to the current,
involving the emission or the absorption of Cooper pairs (charge
$2e$) on the superconducting side.
An incoming electron in the normal side at energy $\mu_S + eV$
drags another electron at energy $\mu_S - eV$, forming a
reflected hole of energy $\mu_S - eV$. The two combined
electrons have a total energy $2 \mu_S$ which corresponds to a
Cooper pair, and are thus allowed to be transfered to the
superconducting side. Now the above argument for an
oscillatory time dependence can be repeated, since the incoming
electron wave function oscillates as $\psi_e \sim \exp[-i (\mu_S
+ eV)t/\hbar]$, whereas the hole wave function oscillates as
$\psi_h \sim \exp[-i (\mu_S - eV)t/\hbar]$. The noise
combines these dependences in the product
$\psi_e \psi_h^*$ which now oscillates at 
the Josephson frequency $2eV/\hbar$ corresponding to the
singularity.  In a junction containing a single 
superconductor, the singularity therefore appears
in the noise rather than the current, and the detection of
this frequency can be considered as an analog
to the Josephson effect.

\subsection{Large biases}
\label{para_large_bias}
The applied bias is now larger than (or comparable to) the
gap, so that the scattering matrix elements depends on the
energy. A specific description is needed to characterize this 
dependence: the BTK model \cite{BTK} is particularly suited for
this purpose as it allows a description with a minimal
number of parameters.

A local tunnel barrier $V_B(x)=V_B\delta(x)$ 
is introduced at the boundary.
Since the complete knowledge of the $S$ matrix of the junction
is necessary to compute the noise, the quasi-particle states in
the superconductor are specified. This only makes
sense if these states are not evanescent ($E>\Delta$). 
The Bogolubov-de Gennes equations for these states are solved
on the superconducting side.
Using the continuity of the wave functions
at the interface, and specifying the discontinuity of
their derivative, one obtains the $S$ matrix
elements (Appendix \ref{S_matrix_BTK}).

\begin{figure}[htb]
\begin{center}
\mbox{\epsfig{file=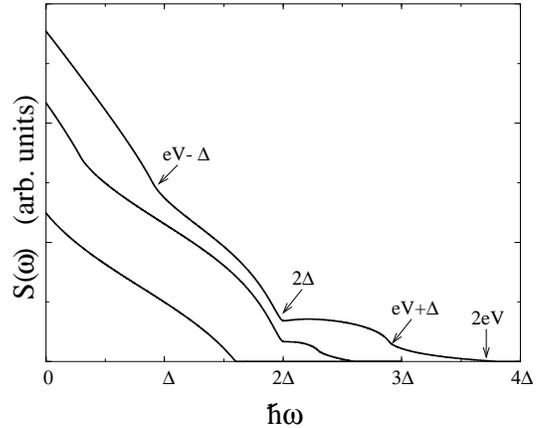,width=8.5cm}}
\narrowtext
\caption[to]{
Noise in a N-S junction as a function of frequency,
with intermediate barrier transparency ($Z=1$),
for several values of the applied bias :
$eV = 0.8 \Delta$, $eV = 1.3 \Delta$, $eV = 1.9 \Delta$.
}
\label{sing1}
\end{center}
\end{figure}

\begin{figure}[htb]
\begin{center}
\mbox{\epsfig{file=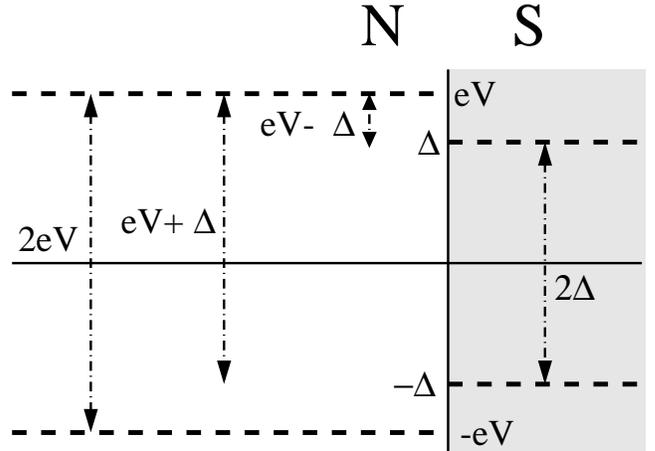,width=8.5cm}} 
\narrowtext
\caption[to]{
Energy diagram when the bias is larger than the gap.
Relevant intervals
associated with the cusps/singularities are shown.
}
\label{sing2}
\end{center}
\end{figure}

\begin{figure}[htb]
\begin{center}
\mbox{\epsfig{file=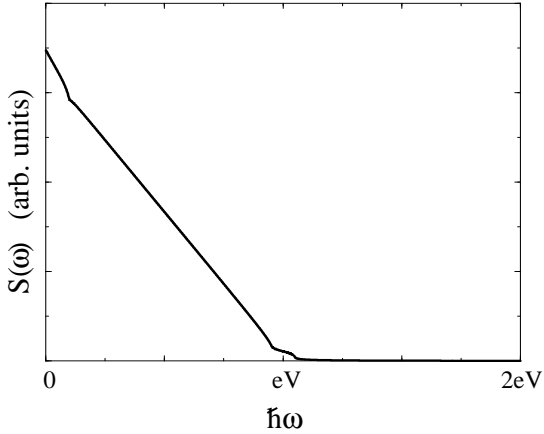,width=8.5cm}} 
\narrowtext
\caption[to]{
Noise in a N-S junction as a function of frequency,
with intermediate barrier transparency ($Z=1$),
for a large value of the bias ($eV = 20 \Delta$).
}
\label{sing3}
\end{center}
\end{figure}

The energy integrals in Eqs. (\ref{eq_int_bruit_1}) and
(\ref{eq_int_bruit_2}) are performed numerically.
Plotting the noise as a function of frequency,
additional cusps or singularities are found at
$\omega=(eV-\Delta)/\hbar$, $\omega=(2\Delta)/\hbar$,
$\omega=(eV+\Delta)/\hbar$, on top of the Josephson singularity
at $\omega=2 eV /\hbar$ (figure \ref{sing1}). All
these frequencies can be illustrated on an energy diagram
(figure \ref{sing2}). This numerical
calculation can also be performed for small biases, yielding
full agreement with the
previous calculation (\ref{eq_bruit_NS_simple}). Another
interesting limit arises when $eV \gg \Delta$. In this case,
transport is dominated by single quasi-particle transfer 
with a charge $e$, whereas the 
contribution of Andreev reflection is small. Thus
similar results to those of normal-normal metals
junction are expected. This is obviously the case (figure
\ref{sing3}), even though the above mentioned singularities
can still be identified.
One may object that if the applied bias is too large, the
non-equilibrium processes dominate, and the previous assumptions
are not correct anymore because of heating effects. This limit
is then valid for superconductors with a small gap
($\Delta/k_B\sim 0.1 K$) because the condition $eV \gg
\Delta$ may be satisfied in a near-equilibrium situation.

In order to visualize the additional singularities,
the argument invoking the oscillatory time dependence 
can be used once again. This time, because of the large 
value of the bias, several charge transfer processes occur:
a)
Andreev reflection is still there, and it implies the same
singularity at the Josephson frequency $2eV/\hbar$.
b)
Electrons in the normal side are transmitted as electron-like quasi-particles
in the superconductor. Wave functions oscillate as
$\psi_{N,e} \sim \exp[-i (\mu_S + eV)t/\hbar]$ and
$\psi_{S,e} \sim \exp[-i (\mu_S + \Delta)t/\hbar]$,
and the overlap gives a singularity at $(eV-\Delta)/\hbar$.
Note that the same transfer process occurs with holes and hole-like
quasi-particles, giving the same singularity.
c)
Electrons in the normal side are transmitted as hole-like quasi-particles
in the superconductor (Andreev transmission).
Here, the time dependence of the wave functions is
$\psi_{N,e} \sim \exp[-i (\mu_S + eV)t/\hbar]$ and
$\psi_{S,h} \sim \exp[-i (\mu_S - \Delta)t/\hbar]$,
implying a singularity at $(eV+\Delta)/\hbar$.
The same transfer process exists with holes and electron-like
quasi-particles.
d)
Andreev reflection also occurs on the superconducting side,
when electron-like quasi-particles are reflected as hole-like 
quasi-particles, and vice versa. Wave functions oscillates as
$\psi_{S,e} \sim \exp[-i (\mu_S + \Delta)t/\hbar]$ and
$\psi_{S,h} \sim \exp[-i (\mu_S - \Delta)t/\hbar]$,
giving a singularity at the frequency $2 \Delta/\hbar$.
These four singularities are summarized in Fig. \ref{sing1}.

\subsection{Non-stationary Aharonov-Bohm effect}
\label{para_NSAB}
Finite frequency measurements can represent a real challenge for
noise, so it is interesting to imagine a scenario where an
alternating field superposed to the DC bias allows to probe the
finite frequency effect.
The non-stationary Aharonov--Bohm effect has been
introduced several years ago in a normal conductor connected
to reservoirs \cite{Lesovik Levitov}. In this proposal, a time
dependent vector potential is applied in a confined region
$[x_1,x_2]$ of the conductor, which adds a phase to the
electrons and holes. The phase is chosen to be a periodic
function of time $\Phi(t)=\Phi_a\sin(\Omega t)$ with
$\Phi_a\equiv 2\pi\int_{x_1}^{x_2} dx A_x/\phi_0$,
and where $\phi_0=hc/e$ is the normal flux quantum.
The most striking consequence is the presence of steps in
the derivative of the shot noise with respect to
the voltage $\partial S/\partial eV$
when the applied bias $eV$ is a multiple of the frequency
of the perturbation $\hbar \Omega$.
Moreover, the gaps between the steps are
non-monotonic with the amplitude of the harmonic 
vector potential. This effect has been experimentally  observed
in normal diffusive samples \cite{Schoelkopf2}.

\begin{figure}[htb]
\begin{center}
\mbox{\epsfig{file=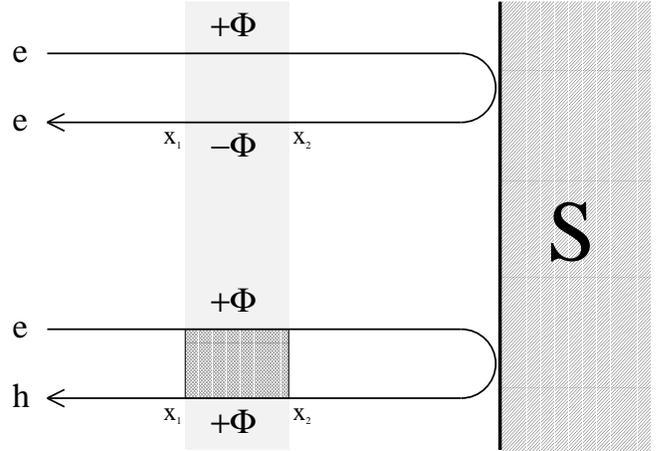,width=8.5cm}} 
\narrowtext
\caption[to]{
The NS boundary. The light
shaded region determines where the electron/hole wave function
may accumulate phase: schematic description of the two 
scattering processes, Andreev and normal reflection.
}
\label{AndreevNS}
\end{center}
\end{figure}

Here, this result is extended to an N-S junction
using the same framework (figure \ref{AndreevNS}). The perturbation
remains confined in the interval $[x_1,x_2]$ near the boundary
and is assumed to contain an adiabatic time modulation.
The main difference with the previous case is due
to the Andreev reflection. The wave function of an incoming electron
accumulates a phase $\Phi(t)$ in the region where the potential
is confined, but after a normal reflection the same phase is
subtracted. On the contrary, if the electron is Andreev
reflected, the outgoing hole accumulates a phase $\Phi$,
totalizing a phase $2 \Phi$ for the complete reflection process.
So in the Andreev regime,
the $S$ matrix of the junction can be written as follow:
\begin{equation}
S=
\left(
\begin{array}{cc}
s_{ee} & s_{eh} \, e^{-2i\Phi(t)}
\\
s_{he} \, e^{2i\Phi(t)} & s_{hh}
\end{array}
\right)~,
\end{equation}
where $s_{ee}$, $s_{eh}$, $s_{he}$ and $s_{hh}$ are the standard 
matrix elements describing the N-S boundary only (without the
external time perturbation). %
In contrast to the usual Aharonov-Bohm effect, no closed topology
is imposed: the flux does not have to be enclosed in a loop,
and the current is {\it not} periodic in the accumulated phase 
$2\Phi(t)$.
Moreover, the effect of the perturbation on the average 
current is straightforward in the limit where the
probability of Andreev reflection $R_A$ depends
weakly on the energy: it brings a periodic modulation of 
the current $\Delta I=(4e^2/h)R_A[\hbar\Omega/e]\Phi_a\cos(\Omega t)$.
The most striking consequence of the flux occurs when
the current-current correlations are considered: 
the modulation leads to a non-monotonic effect as a 
function of phase, in contrast with the 
electromotive force action on the current.

\begin{figure}[htb]
\begin{center}
\mbox{\epsfig{file=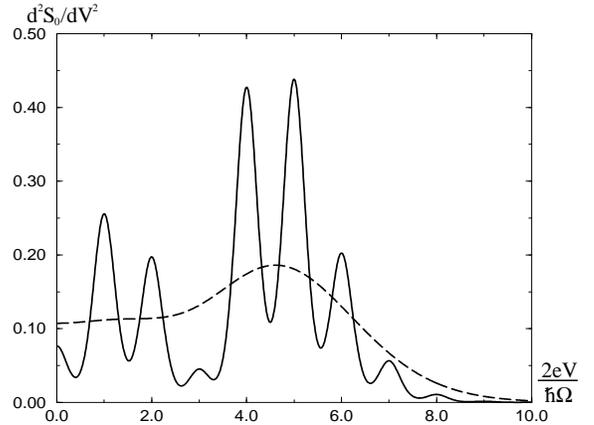,width=8.5cm}} 
\narrowtext
\caption[to]{
Non-stationary Aharonov-Bohm effect: plot of 
$\partial^2S/\partial V^2$, expressed in units of 
$(8e^4/\pi\hbar^2\Omega)R_A(1-R_A)$, as a function of $2eV/\hbar\Omega$,
with the choice $\Phi_a=3$.
For $2k_B T=0.2\hbar\Omega$ (full line) and for 
$2k_B T=\hbar\Omega$ (dashed line).
}
\label{NSABE}
\end{center}
\end{figure}

However, because of this periodic perturbation, translational
invariance in time is broken, and the process becomes
non-stationary. So the noise depends in general on two
frequencies and thus can be written as:
\begin{eqnarray}
\nonumber
\widetilde{S}(\Omega_1,\Omega_2) &=&
\int \int dt_1 dt_2 e^{i \left( \Omega_1 t_1 + \Omega_2 t_2 \right)}
\\
&& \hspace{1.5cm}
\Bigl( \langle  I(t_1) I(t_2) \rangle - \langle I \rangle^2 \Bigr)
~.
\end{eqnarray}
This double Fourier transform can be rewritten as:
\begin{equation}
\nonumber
\widetilde{S}(\Omega_1,\Omega_2) =
\sum_{m=-\infty}^{+\infty}  
2\pi \delta(\Omega_1+\Omega_2-m\Omega) S^{(m)} (\Omega_2)
~.
\end{equation}
The zero harmonic, which is proportional to
$\delta(\Omega_1+\Omega_2)$ is the most standard quantity to
study. From Eq. (\ref{eq_corr_temp}), its zero-frequency
expression is given by:
\begin{eqnarray}
\nonumber
S^{(0)} (0) &=&
\frac{e^2 \hbar^2}{2m^2 v_F^2} \frac{1}{\left( 2 \pi \hbar \right)^2}
\int_{-\infty}^{+\infty} \!\!\! dt \int_{-\infty}^{+\infty} \!\!\! dt'
\\
\nonumber
&& \hspace{5mm} \times
\int_0^{+\infty} \!\!\! dE \int_0^{+\infty} \!\!\! dE' \sum_{m,n} \Biggl\{
\\
\nonumber
&& \hspace{5mm}
f_m(E)(1-f_n(E')) e^{i(E'-E)(t'-t)/\hbar}
\\
\nonumber
&& \hspace{5mm}
\times \Bigl[
  A_{N m N n}(E,E',t)   A^*_{N m N n}(E,E',t')
\\
\nonumber
&& \hspace{5mm}
+ B^*_{N m N n}(E,E',t) B_{N m N n}(E,E',t')
\\
\nonumber
&& \hspace{5mm}
+ A_{N m N n}(E,E',t)   B_{N m N n}(E,E',t')
\\
\nonumber
&& \hspace{5mm}
+ B^*_{N m N n}(E,E',t) A^*_{N m N n}(E,E',t')
\Bigr]
\\
\nonumber
&& \hspace{5mm}
+f_m(E) f_n(E') e^{-i(E+E')(t'-t)/\hbar}
\\
\nonumber
&& \hspace{1cm}
\times
C^*_{N m N n}(E,E') 
\Bigl( C_{N n N m}(E',E,t')
\\
\nonumber
&& \hspace{2cm}
 + C_{N m N n}(E,E',t') \Bigr)
\\
\nonumber
&& \hspace{5mm}
+(1-f_m(E)) (1-f_n(E')) e^{i(E+E')(t'-t)/\hbar}
\\
\nonumber
&& \hspace{5mm}
\times  \Bigl( C_{N n N m}(E',E,t) + C_{N m N n}(E,E',t) \Bigr)
\\
&& \hspace{1cm}
\times
C^*_{N m N n}(E,E',t')
\Biggr\}
~.
\end{eqnarray}
Performing this integral at finite temperature
from the explicit time dependence of the current 
matrix elements and using the generating function of the Bessel
functions $J_n$, one obtains:
\begin{eqnarray}
\nonumber
S^{(0)}(0) &=&
\frac{4e^2}{h} R_A (1- R_A) \sum_{m=-\infty}^{+\infty} J_m^2(2\Phi_a) F_V( m \hbar \Omega)
\\
&&
+\frac{8e^2}{h} R_A^2 k_B T 
~.
\label{Noise AB}
\end{eqnarray}
where the temperature dependence appears in the form:
\begin{equation}
F_V(m \hbar \Omega)= (2eV - m \hbar \Omega)\coth[(2eV-m \hbar \Omega)/2k_BT]~.
\end{equation}
Note the factor $2$ reminiscent of the Cooper pair charge 
in the argument of the Bessel function,
which originates from the accumulated phase $2\phi$.
Now the derivative of the noise
with respect to the voltage is taken:
\begin{equation}
\frac{\partial S^{(0)}(0)}{\partial V} \simeq
\frac{8e^3}{h} R_A (1- R_A) \sum_{m=-M}^{+M} J_m^2(2\Phi_a)
~.
\label{eq_noise_derivative}
\end{equation}
$F_V(m \hbar \Omega)$ specifies how the steps in the noise
derivative are smeared with temperature. In Eq. (\ref{eq_noise_derivative}) 
the sum over harmonics has a cutoff at 
$M = \lfloor 2eV/\hbar \Omega \rfloor$. 
In experiments, \cite{Schoelkopf2} it is more convenient 
to characterize the non-monotonic dependence on voltage by 
taking the second derivative of the Aharonov-Bohm contribution 
to the noise. This is illustrated
for two distinct temperatures in figure \ref{NSABE}.
For small temperatures
$k_B T < \hbar\Omega/2$, the noise steps are individually
resolved, and one observes oscillations as a function of
$2eV/\hbar\Omega$, with a clustering of peaks with a large amplitude. 
For larger temperatures ($k_B T > \hbar\Omega/2$), one expects the 
signal to vanish but this is obviously not the case:
although individual peaks separated by $\hbar\Omega$
can no longer be identified, clusters of peaks (or
clusters of ``large'' steps in the noise derivative)
continue to give an average contribution  to the non-stationary 
AB effect. This robustness enhances the likelihood
of experimental observations.

The corresponding experiment was successfully
recently achieved in diffusive conductors \cite{Kozhevnikov}.
For comparison with theory, the current spectral
density is averaged over the transmission channels, and
Eq. (\ref{Noise AB}) becomes:
\begin{eqnarray}
\nonumber
S^{(0)}(0) &=& 4 k_B ~ T ~ G_{NS}(1-\eta) \\
&&
 + 2 \eta ~ G_{NS}
\sum_{m=-\infty}^{+\infty} J_m^2(2\Phi_a) F_V( m \hbar \Omega)
~,
\end{eqnarray}
where $G_{NS}$ is the differential conductance of the device,
and $\eta=1/3$ the suppression factor for a normal diffusive conductor.
The second derivative with respect to the voltage of the measured noise
clearly shows peaks at the Josephson frequency. This constitutes
a rather robust experimental check of the presence of an
effective charge $2e$ in the fluctuation spectrum of single NS
junctions. 
\section{Hanbury-Brown and Twiss gedanken experiment with a
superconductor} \label{para_HBT}
\subsection{Introduction}
So far, attention on effective charges have been the primary
focus. Effects associated with the statistics of the charge
carriers are now considered. In the mid-1950s, Hanbury-Brown
and Twiss \cite{HBT} described a new type of interferometer in
order to find the size of a radio star by measuring the
correlations between the signals of two aerials. This experiment
was followed by another one using a coherent light source, where
a mercury arc lamp beam was
partitioned by a splitter into a reflected and a transmitted
part. Intensity correlations between reflected and transmitted
beams were found to be positive. This result can be explained by
the quantum statistical properties of photons, which are bosons.
Particles in a beam of bosons tend to cluster together
(bunching), so the probability to detect simultaneously two
photons (one in each beam) is non-zero, and therefore
correlations are positive. 
On the other hand, two
indistinguishable fermions exclude each other because of the
Pauli principle (anti-bunching), and consequently, reflected and
transmitted beams are anti-correlated \cite{Martin
Landauer,Buttiker}.

More recently, two analogs to the original Hanbury-Brown
and Twiss experiment, using
fermions propagating in semiconductors (electrons) -- instead of
photons is vacuum -- were achieved. Negative correlations
were expected, and
experimentally verified \cite{Henny,Oliver}. Here
an hybrid system is envisioned: a superconductor is introduced
just next to the beam splitter. This implies that 
transport involves Cooper pairs on the superconducting side,
which have a finite penetration length on the normal side,
the essence of the proximity effect 
\cite{Pannetier,Petrashov,Van Wees}.
While Cooper pairs are not strictly bosons,
an arbitrary number of these can exist in the same momentum state,
so bosonic statistics could be detected in such a system. 
Thus the possibility for {\it positive} noise correlations
cannot be ruled out.
Below, it is shown that the sign of the noise correlations
depends on the transparency of the beam splitter.
\subsection{Model}
\begin{figure}[htb]
\begin{center}
\mbox{\epsfig{file=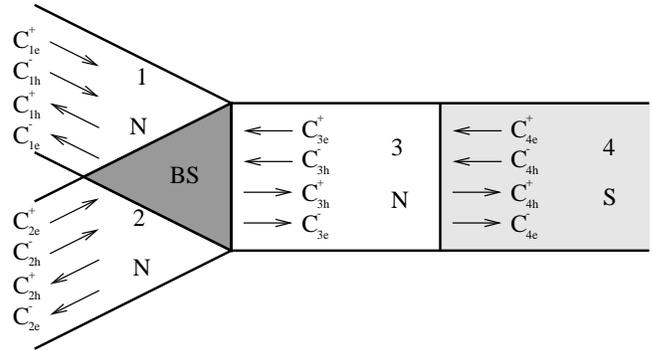,width=8.5cm}} 
\narrowtext
\caption[to]{
Two normal leads (1 and 2) are linked by a semi-transparent
beam splitter (BS) and connected
to a superconductor (4) via a normal region (3).
}
\label{jonctionY}
\end{center}
\end{figure}

The device is composed of two normal leads (1
and 2, see figure \ref{jonctionY}) linked by a
semi-transparent beam splitter (BS) and connected to a
superconductor. 
The state corresponding to an electron incoming
(outgoing) in (from) the lead $i$ is labelled
$c_{i \, e}^+$ ($c_{i \, e}^-$). The hole incoming
(outgoing) in (from) the lead $i$ is described by
$c_{i \, h}^-$ ($c_{i \, h}^+$) (see figure \ref{jonctionY}).
With this convention, the $S$ matrix
of the whole system is defined as:
\begin{equation}
\left(
\begin{array}{c}
c_{1e}^- \\ c_{1h}^+ \\ c_{2e}^- \\ c_{2h}^+ \\ c_{4e}^- \\ c_{4h}^+
\end{array}
\right)
= S
\left(
\begin{array}{c}
c_{1e}^+ \\ c_{1h}^- \\ c_{2e}^+ \\ c_{2h}^- \\ c_{4e}^+ \\ c_{4h}^-
\end{array}
\right)
~.
\end{equation}
The goal is now to calculate
the expression of the noise correlations between leads 1 and 2.
From Eq. (\ref{eq_corr_bruit}) and in the limit of zero
temperature, it is possible to show
that the expression for the noise correlations reduces to:
\begin{eqnarray}
\nonumber
S_{12}(0)&=& \frac{2e^2}{h}\int_0^{eV} dE \Bigl[
\sum_{i,j=1,2}
  \left( s^*_{1 i ee} s_{1 j eh} - s^*_{1 i he} s_{1 j hh} \right)
\\
\nonumber
&& \hspace{1.5cm}
\times  \left( s^*_{2 j eh} s_{2 i ee} - s^*_{2 j hh} s_{2 i he} \right)
\\
\nonumber
&& \hspace{1cm}
+ \sum_{i=1,2 \, ; \, \alpha=e,h}
  \left( s^*_{1 i ee} s_{14e\alpha} - s^*_{1 i he} s_{14h\alpha} \right)
\\
\label{eq_corr_Y}
&& \hspace{1.5cm}
\times  \left( s^*_{24e\alpha} s_{2 i ee} - s^*_{24h\alpha} s_{2 i he} \right)
\Bigr]
~.
\end{eqnarray}
The sign of Eq. (\ref{eq_corr_Y}) cannot be determined at this
stage: it depends on the specific form of the $S$ matrix.
Thus, for analytical purposes, a simple model is proposed, where
the beam splitter is dissociated from the N-S junction (see
figure \ref{jonctionY}). To establish the $S$ matrix
of the junction, one needs to combine the $S$
matrices of the beam splitter and of the N-S junction (between
3 and 4).
\subsection{$S$ matrix of the splitter}
The beam splitter is described by a $S$ matrix $S_M$
which gives the outgoing states as a function of incoming ones.
For electrons one obtains:
\begin{equation}
\left(
\begin{array}{c}
c_{1e}^- \\ c_{2e}^- \\ c_{3e}^-
\end{array}
\right)
= S_{M_e}
\left(
\begin{array}{c}
c_{1e}^+ \\ c_{2e}^+ \\ c_{3e}^+
\end{array}
\right)
~.
\end{equation}
Expression of $S_{M_e}$ is identical to that of Ref.
\cite{Gefen}: %
\begin{equation}
S_{M_e} =
\left(
\begin{array}{ccc}
a & b & \sqrt{\varepsilon}
\\
b & a & \sqrt{\varepsilon}
\\
\sqrt{\varepsilon} & \sqrt{\varepsilon} & -(a+b)
\end{array}
\right)
~,
\end{equation}
where $a = \left( \sqrt{1-2\varepsilon} -1 \right)/2$,
$b = \left( \sqrt{1-2\varepsilon} +1 \right)/2$, and
$\varepsilon$ can vary from 0 to $1/2$.
$S_{M_e}$ depends only on a single parameter $\varepsilon$
which monitors the transparency of the splitter.
For example if $\varepsilon = 0$ no transmission occurs from
region (1) or (2) to region (3).
A similar relation holds for holes:
\begin{equation}
\left(
\begin{array}{c}
c_{1h}^+ \\ c_{2h}^+ \\ c_{3h}^+
\end{array}
\right)
= S_{M_h}
\left(
\begin{array}{c}
c_{1h}^- \\ c_{2h}^- \\ c_{3h}^-
\end{array}
\right)
~.
\end{equation}
Note that the splitter does not couple electrons and holes.
Expression of the matrix for holes is given by the relation
$S_{M_h}(E) = S^*_{M_e}(-E)$, as no magnetic field is assumed
to be present here,  but since $S_{M_e}$ is real and does not
depend on the energy $S_{M_h} = S_{M_e}$.
\subsection{Small biases: Andreev regime}
\begin{figure}
\begin{center}
\mbox{\epsfig{file=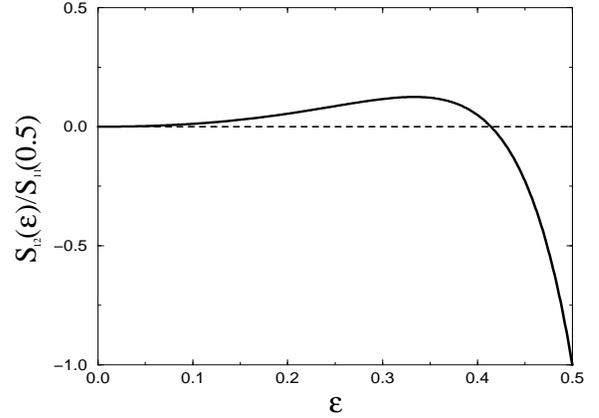,width=8.5cm}} 
\narrowtext
\caption[to]{
Noise correlations between the two normal reservoirs
of the device, as a function of transmission of the 
beam splitter, showing both positive and negative correlations.
}
\label{pos_corr1}
\end{center}
\end{figure}

When the applied bias is much
smaller than the gap ($eV \ll \Delta$), 
Andreev reflection between 3 and 4 is the only transmission
process. One can then write
\cite{Beenakker2}:
\begin{equation}
\left(
\begin{array}{c}
c_{3e}^+ \\ c_{3h}^-
\end{array}
\right)
=
\left(
\begin{array}{cc}
0 & \gamma \\
\gamma & 0
\end{array}
\right)
\left(
\begin{array}{c}
c_{3e}^- \\ c_{3h}^+
\end{array}
\right)
~,
\end{equation}
with $\gamma = e^{-i \arccos (E/\Delta)}$. In such a case, all matrix
elements like  $s_{14pq}$ or $s_{24pq}$ (with $p,q=e,h$) are
zero. Setting $x=\sqrt{1-2\varepsilon}$ it is possible to show
that:
\begin{eqnarray}
s_{11ee} \! = \! s_{11hh} \! = \! s_{22ee} \! = \! s_{22hh} \! &=& \! \frac{(x-1)(1+\gamma^2 x)}{2(1-\gamma^2 x^2)} ~,
\\
s_{21ee} \! = \! s_{21hh} \! = \! s_{12ee} \! = \! s_{12hh} \! &=& \! \frac{(x+1)(1-\gamma^2 x)}{2(1-\gamma^2 x^2)} ~,
\end{eqnarray}
\begin{eqnarray}
\nonumber
s_{11eh} \! = \! s_{21eh} \! &=& \! s_{12eh} \! = \! s_{22eh} \! = \! s_{11he} \! = \! s_{21he} \! = \! s_{12he} \! = \! s_{22he} \!
\\
 &=& \! \frac{\gamma (1-x)(1+x)}{2(1-\gamma^2 x^2)} ~.
\end{eqnarray}
Because $E \ll \Delta$ one can make the assumption that
$\gamma \simeq -i$. Since the $S$ matrix elements do not
depend on the energy anymore, the integral (\ref{eq_corr_Y}) can
be performed. One finally obtains:
\begin{equation}
\label{eq_correlations_splitter}
S_{12}(0) = \frac{2e^2}{h} eV \frac{\varepsilon^2}{2(1-\varepsilon)^4}
\left( -\varepsilon^2 -2\varepsilon+1 \right)~.
\end{equation}
The noise correlations vanish at $\varepsilon=0$, when 
conductors $1$ and $2$ are equivalent to a two--terminal device
decoupled from the  superconductor, and in addition, $S_{12}$
vanishes when  $\varepsilon=\sqrt{2}-1$. A plot of $S_{12}$
(normalized to the noise in $1$ (or $2$) at $\varepsilon=1/2$)
as a function of 
the beam splitter transmission (figure \ref{pos_corr1}) indicates 
that indeed, the correlations are positive (bosonic) for 
$0<\varepsilon<\sqrt{2}-1$ and negative (fermionic) for 
$\sqrt{2}-1<\epsilon<1/2$. At maximal transmission into the 
normal leads ($\epsilon=1/2$), the correlations
give the negative minimal 
value: electrons and holes do not interfere and propagate
independently into the normal terminals. 
This is the signature of a purely fermionic system. 
When the transmission $\epsilon$ is decreased, Cooper 
pairs may leak in region $3$ because of 
multiple Andreev processes at the boundary.
Further reducing the 
beam splitter transmission allows to balance the contribution 
of Cooper pairs with that of normal particles. 
Expression (\ref{eq_correlations_splitter}) predicts maximal (positive)
correlations at $\epsilon=1/3$: a compromise
between a high density of Cooper pairs and weak transmission.  

\subsection{Larger biases}
If the applied bias is
greater than the gap, transmission of quasi-particles
between 3 and 4 is now allowed,
and one has to take into account the energy dependence
of the $S$ matrix elements. 
As in section \ref{para_large_bias}, the BTK model is chosen,
and thus quasi-particles at arbitrary energy can be handled. For
simplicity, the same beam splitter is still used,
assuming that its $S$ matrix is independent of the energy.
Even though this may appear
as a crude approximation, this remains correct for example when
the superconductor has a small gap. The $S$ matrix of the whole
system is computed in appendix \ref{S_matrix_Y}.

\begin{figure}[htb]
\begin{center}
\mbox{\epsfig{file=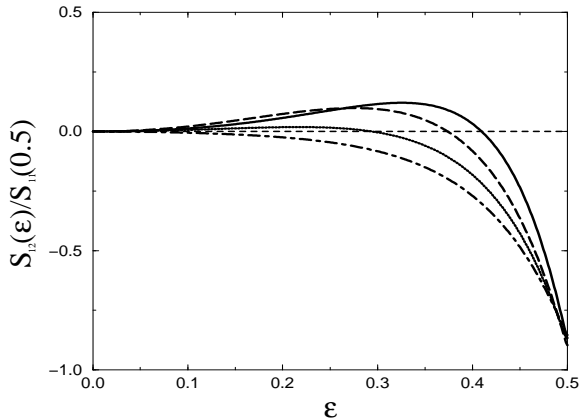,width=8.5cm}} 
\narrowtext
\caption[to]{
Noise correlations using an N-S boundary modeled by BTK
for a high transparency barrier ($Z=0.1$). From top
to bottom, $eV/\Delta=0.5,0.95,1.2,1.8$.
}
\label{pos_corr2}
\end{center}
\end{figure}

A numerical calculation of noise correlations is performed
with the help of Eq. (\ref{eq_corr_Y}). If one considers a
high transparency barrier ($Z=0.1$) and a small bias, one finds a
good agreement with previous analytical results (figure
\ref{pos_corr2}), except that the noise correlations (still
normalized to the noise in a normal lead when $\varepsilon=1/2$)
do not quite reach the minimal value at $\varepsilon=1/2$. This
is an early signature of the potential barrier at the NS
interface. The more the bias is increased, the  weaker are the
positive correlations. If the voltage is large enough (beyond
the gap) positive correlations are completely destroyed.
As in the analogy with the normal-normal junction encountered
previously, Cooper pairs contribute only for a small part to
the transport, and the system loses its bosonic features.

\begin{figure}[htb]
\begin{center}
\mbox{\epsfig{file=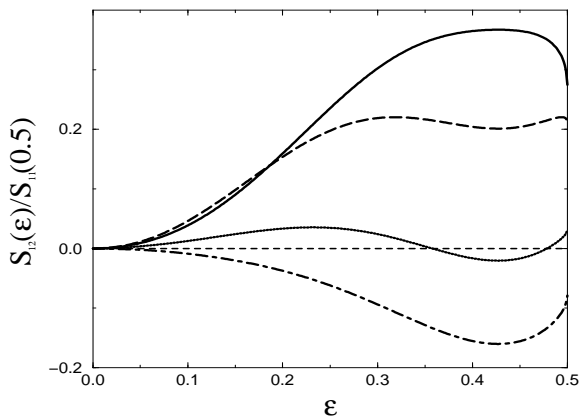,width=8.5cm}} 
\narrowtext
\caption[to]{
Noise correlations using an N-S boundary modeled by BTK
for intermediate transparency:
$Z=1$ (same biases as in Fig. \ref{pos_corr2}).
}
\label{pos_corr3}
\end{center}
\end{figure}

Intermediate transparencies are now considered
($Z=1$, figure \ref{pos_corr3}), and a strikingly different
behavior is obtained. For weak biases, noise correlations remain
positive over the whole range of $\varepsilon$.
It is possible to find an appropriate value (for example $eV=0.95 \Delta$)
in order to observe oscillations between positive and negative
values of the correlations.
Further increasing bias, correlations again become negative over
the whole range of $\varepsilon$. Calculations for larger values of 
$Z$ confirm the tendency of the system towards
dominant positive correlations at low biases
with $S_{12}(\epsilon)/S_1(1/2)> 1$ over a wide range of 
$\epsilon$ (not shown). 
The phenomenon of positive correlations in fermionic 
systems with a superconducting injector is thus 
{\it enhanced} by the barrier opacity at the N-S boundary. 
Nevertheless, at the same time, for opaque barriers,
the absolute magnitude of $S_{1}$ and $S_{12}$
becomes rather small, which limits the possibility of an
experimental check in this regime.

\begin{figure}[htb]
\begin{center}
\mbox{\epsfig{file=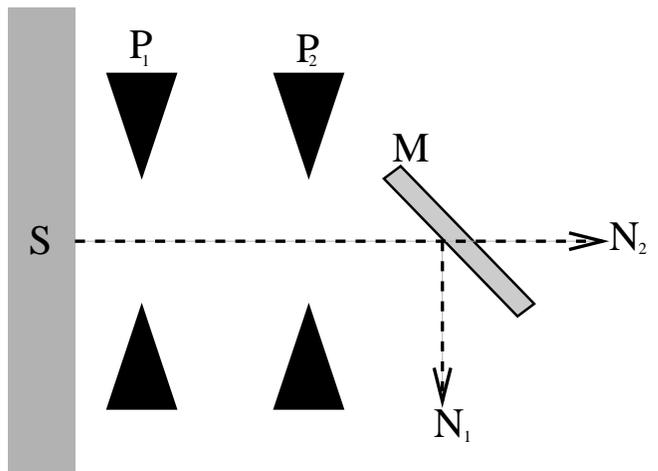,width=8.5cm}} 
\narrowtext
\caption[to]{
Proposed device for the 
observation of positive/negative correlations; 
at the boundary of a superconductor ($S$), two point 
contacts ($P_1$) and ($P_2$) are connected to a 
semi-transparent mirror ($M$).
}
\label{fig_device}
\end{center}
\end{figure}

A suggestion for the experimental device is depicted 
figure \ref{fig_device}. Assume that a 
high mobility two dimensional electron gas has
a rather clean interface with a superconductor
\cite{Dimoulas}. 
A first point contact ($P_1$) close to the 
interface selects a maximally occupied electron channel.
The beam of electrons is incident on a semi--transparent mirror similar
to the one used in the Hanbury--Brown and Twiss
fermion analogs \cite{Henny,Oliver}. A second point 
contact located in front of the mirror ($P_2$) allows 
to modulate the reflection of the splitter in order 
to monitor both bosonic and fermionic noise correlations. 
In addition, by choosing a superconductor with 
a relatively small gap, 
one could observe the dependence of the
correlations on the voltage bias without encountering 
heating effects in the normal metal. 

An alternative interpretation of these positive correlation has
been proposed \cite{Blanter_Buttiker,Gramespacher} in terms of the sign of
the effective charges obtained from the spectral weight
associated to electrons and holes. Nevertheless, statistical
analogies are often useful in condensed matter physics as these
allow to isolate the dominant  behavior in the transport
characteristic of a given system.  For instance, in the
fractional quantum Hall effect, dissipationless transport
occurs because electrons with an odd number of flux quanta
represent a composite boson \cite{MacDonald}.

\section{Conclusion}
Both dynamical and statistical aspects of noise have been
presented in a unified formalism. Expressions for the
finite-frequency noise and the noise correlations for a
conductor containing an 
arbitrary number of terminals, connected to a superconductor (Eq.
(\ref{eq_bruit}) and (\ref{eq_corr_bruit})) have been derived. In
a first step, a single N-S junction was considered. In the
Andreev regime, the noise spectral density presents a singularity
at the Josephson frequency, and vanishes beyond. This can be
interpreted in terms of an effective charge $2e$ transfered at
the boundary. Note that this argument fails if the
bias voltage is increased above the gap. In this case, both
a single and a double charge transfer are allowed. Thus the
effective charges which are identified in the spectral noise
density plots are not well defined ($e<e^*<2e$). 

The non-stationary Aharonov-Bohm effect has been proposed as a
tool to analyze these features in a zero frequency measurement.
It allows the observation
of peaks in the second derivative of the noise with respect to
the voltage when the frequency of the applied perturbation is
commensurate with the Josephson frequency. Because these kinds of
measurements are in principle easier to achieve, this
non-stationary Aharonov-Bohm effect has caught the interest of
experimentalists \cite{Schoelkopf2,Kozhevnikov}, who have
provided a confirmation of the theoretical
predictions of this paper. Thus, a single N-S junction is an
adequate system to observe the Josephson frequency with only one
superconductor instead of two, as in the usual Josephson effect. 

In a second step, the feasibility of an analog to the fermionic
Hanbury--Brown and Twiss  experiment
with a superconductor has been addressed. Correlations are shown
to be either positive or negative, depending on the reflection
coefficient of the beam  splitter. Therefore, such a {\it
fermionic} system can exhibit a {\it bosonic} behavior. 
A qualitative interpretation of this puzzling result 
is reached as follows: when the transmission at the interface
decreases,   Cooper pairs leak on the normal side, and these can
be considered as ``composite bosons'', hence the positive
statistical signature.  Recent experiments in the ``normal''
fermionic Hanbury--Brown and Twiss analog \cite{Henny,Oliver}
were performed successfully, and these experiments could
possibly be extended to the case presented here, giving the
opportunity to observe for the first time positive correlations
in a fermionic system.

Nevertheless, the issues of this paper present some limitations.
All the calculations have been performed in the single channel
case, even if a generalization to several channels is possible.
However, this extension should not bring major changes. The
assumption that the $S$ matrix elements are independent on
the energy is correct as long as the applied bias remains small
enough. For larger biases, the BTK model takes this energy
dependence into account, but only allows numerical calculations.
Moreover, further increasing the bias leads to a non-equilibrium
situation which cannot be described with this formalism. An
appropriate approach would be to employ the Keldysh Green's
functions method. However, in order to go further than the
present calculations, heating effects should be taken into
account self-consistently. 
Nevertheless, the physics presented here is rather robust,
and different results, especially concerning the effective charge
$2e$ in the Andreev regime are
not expected. The calculations have been performed for a {\it
arbitrary} $S$-matrix describing a specific sample, 
but as pointed out in Ref. \cite{Kozhevnikov}, all these
results can be extended to the diffusive case without
difficulty by averaging over the transmission channels
with the standard methods of Ref. \cite{Beenakker3}.

Future considerations may include the transport characteristics 
of other types of superconductors (high $T_c$ superconductors),
where the gap varies in momentum space. 
Recently, the noise in a junction between a
normal metal and a $d$-wave superconductor has been calculated
\cite{Zhu Ting}. Moreover, as it was emphasized above, the study
of the statistics and of the effective charges of quasi-particles
is a relevant issue in condensed matter systems. For
instance, in the fractional quantum Hall effect, 
Laughlin's quasi-particles are supposed to obey fractional
statistics \cite{Haldane}. Results concerning 
particles which obey exclusion statistics \cite{Haldane2}
showed that the shot-thermal noise crossover deviates from the 
fermion case \cite{Isakov Martin Ouvry}. 
Another example is the study of the 
transport properties of ``atomic'' composite fermions/bosons,
such as alkali atoms or $^3\mbox{He}$ and $^4\mbox{He}$
which off-equilibrium properties are not known at this time.
Finally, noise correlations have been computed here {\it
only} at zero frequency. Is there anything to gain from
the finite frequency spectrum of noise correlations ? 
This quantity combines the dynamical aspects of current
fluctuations with the statistical nature of 
multi-terminal geometries.

\section*{Acknowledgements}
Work of G.B.L. was partly supported by the 
Russian Fund for Basic Research (N:000216617).
Discussions with D.~C.~Glattli,
G.~Montambaux and A.~Kozhevnikov are gratefully acknowledged.
One of us (J.~T.) acknowledges valuable discussions with D.~Quirion.
\appendix
\section{$S$ matrix elements of N-S junction with a barrier}
\label{S_matrix_BTK}
Following the BTK model \cite{BTK}, with the barrier potential
$V_B(x)=V_B\delta(x)$ and the pair potential $\Delta(x)=\Delta \theta(x)$,
the Bogolubov-de Gennes equations are solved
on each side of the junction, giving the states 
in the normal lead and in the superconductor.
The wave functions are continuous at the N-S boundary.
Integrating the Bogolubov-de Gennes equations on each
side of the junction gives another
condition over the derivatives of the wave functions.
Writing these conditions for both particle types
and both sides of the junction, one obtains four
linear systems of equations with four unknowns, which
are the $S$ matrix elements:
\begin{eqnarray}
s_{NNee} &=& s_{SSee} = \displaystyle -\frac{(u_0^2-v_0^2)(Z^2+iZ)}{\gamma}
\end{eqnarray}
\begin{eqnarray}
s_{NNhh} &=& s_{SShh} = \displaystyle -\frac{(u_0^2-v_0^2)(Z^2-iZ)}{\gamma}
\end{eqnarray}
\begin{eqnarray}
s_{NNhe} &=& s_{NNeh} = -s_{SShe} = -s_{SSeh} = \displaystyle \frac{u_0 v_0}{\gamma}
\end{eqnarray}
\begin{eqnarray}
s_{SNee} &=& s_{NSee} = \displaystyle \frac{u_0 \sqrt{u_0^2-v_0^2} (1-iZ)}{\gamma}
\end{eqnarray}
\begin{eqnarray}
s_{SNhh} &=&  s_{NShh} = \displaystyle \frac{u_0 \sqrt{u_0^2-v_0^2} (1+iZ)}{\gamma}
\end{eqnarray}
\begin{eqnarray}
\nonumber
s_{SNhe} &=& -s_{SNeh} = -s_{NShe} =  s_{NSeh}
\\
&=& \displaystyle \frac{i v_0 \sqrt{u_0^2-v_0^2} Z}{\gamma}
~,
\end{eqnarray}
where:
\begin{equation}
u_0^2 = \frac{1}{2} \left( 1 + \frac{(E^2 - 
\Delta^2)^{1/2}}{E} \right) = 1 - v_0^2
~.
\end{equation}
$Z=mV_B/(\hbar^2 k_F)$ is the relative height of the barrier
and $\gamma \equiv u_0^2 + (u_0^2-v_0^2) Z^2$.  
Note that the unitarity of the $S$ matrix is satisfied.
\end{multicols}
\widetext
\section{$S$ matrix elements of a three terminal N-S junction
with a barrier} \label{S_matrix_Y}
To compute the complete $S$ matrix, one searches which
outgoing states are obtained
when a single particle is injected in a given terminal.
For example, injecting an electron in 4
({\it ie} a state $c^+_{4e}$), one obtains reflected waves
($c^-_{4e}$ and $c^+_{4h}$)
and transmitted waves ($c^-_{1e}$, $c^+_{1h}$, $c^-_{2e}$ and $c^+_{2h}$)
(see Fig. \ref{jonctionY}).
One can therefore deduct the corresponding $S$ matrix
elements: $s_{44ee}$, $s_{44he}$
$s_{14ee}$, $s_{14he}$, $s_{24ee}$ and $s_{24he}$.
This operation is made for all the terminals and for both types
of particle. One obtains:
\begin{eqnarray}
\left(
\begin{array}{cc}
s_{44ee} & s_{44eh}
\\
s_{44he} & s_{44hh}
\end{array}
\right)
&=&
S_{SS}
-(a+b) S_{SN} \left[ 1+(a+b) S_{NN} \right]^{-1} S_{NS} ~,
\end{eqnarray}
\begin{eqnarray}
\left(
\begin{array}{cc}
s_{14ee} & s_{14eh}
\\
s_{14he} & s_{14hh}
\end{array}
\right)
&=&
\left(
\begin{array}{cc}
s_{24ee} & s_{24eh}
\\
s_{24he} & s_{24hh}
\end{array}
\right)
=
\sqrt{\varepsilon} \Bigl( S_{NS} -(a+b) S_{NN} 
 \left[ 1+(a+b) S_{NN} \right]^{-1} S_{NS} \Bigr) ~,
\end{eqnarray}
\begin{eqnarray}
\left(
\begin{array}{cc}
s_{41ee} & s_{41eh}
\\
s_{41he} & s_{41hh}
\end{array}
\right)
&=&
\left(
\begin{array}{cc}
s_{42ee} & s_{42eh}
\\
s_{42he} & s_{42hh}
\end{array}
\right)
=
\sqrt{\varepsilon} \, S_{SN} \left[ 1+(a+b) S_{NN} \right]^{-1} ~,
\end{eqnarray}
\begin{eqnarray}
\left(
\begin{array}{c}
s_{11ee}
\\
s_{11he}
\end{array}
\right)
&=&
\left(
\begin{array}{c}
s_{22ee}
\\
s_{22he}
\end{array}
\right)
=
\left[
\left(
\begin{array}{cc}
a & 0
\\
0 & 0
\end{array}
\right)
+
\left(
\begin{array}{cc}
\varepsilon + a(a+b) & 0
\\
0 & \varepsilon
\end{array}
\right)
S_{NN}
\right]
\left[ 1+(a+b) S_{NN} \right]^{-1}
\left(
\begin{array}{c}
1
\\
0
\end{array}
\right) ~,
\end{eqnarray}
\begin{eqnarray}
\left(
\begin{array}{c}
s_{11ee}
\\
s_{11he}
\end{array}
\right)
&=&
\left(
\begin{array}{c}
s_{22ee}
\\
s_{22he}
\end{array}
\right)
=
\left[
\left(
\begin{array}{cc}
0 & 0
\\
0 & a
\end{array}
\right)
+
\left(
\begin{array}{cc}
\varepsilon & 0
\\
0 & \varepsilon + a(a+b)
\end{array}
\right)
S_{NN}
\right]
\left[ 1+(a+b) S_{NN} \right]^{-1}
\left(
\begin{array}{c}
0
\\
1
\end{array}
\right) ~,
\end{eqnarray}
\begin{eqnarray}
\left(
\begin{array}{cc}
s_{12ee} & s_{12eh}
\\
s_{12he} & s_{12hh}
\end{array}
\right)
&=&
\left(
\begin{array}{cc}
s_{21ee} & s_{21eh}
\\
s_{21he} & s_{21hh}
\end{array}
\right)
=
\left(
\begin{array}{cc}
s_{11ee} + 1 & s_{11eh}
\\
s_{11he} & s_{11hh} + 1
\end{array}
\right)
~.
\end{eqnarray}
Here $S_{NS}$, $S_{SN}$, $S_{NN}$ and $S_{SS}$ are $2 \times 2$ matrices
of which elements have been computed in appendix \ref{S_matrix_BTK}.
All the calculations have been made above the gap. But the obtained results are
valid even if the energy is smaller than the gap, using the correct
values for the $S$ matrix of the  N-S junction: $S_{NS}$ and $S_{SN}$ are zero,
and $S_{NN}$ and $S_{SS}$ are given in appendix \ref{S_matrix_BTK}.
In this case, expressions of $S_{11}$, $S_{22}$, $S_{21}$ et $S_{12}$ remain
the same, but $S_{41}$, $S_{42}$, $S_{14}$ and $S_{24}$ are zero,
and $S_{42}=S_{SS}$.
\begin{multicols}{2}

\end{multicols}
\end{document}